\documentclass[5pt,a4]{article}
\usepackage{graphicx} 
\usepackage[a4paper, total={6in, 8in}]{geometry}
\usepackage[title]{appendix}
\usepackage{amsfonts}
\usepackage{amsopn}
\usepackage{amsmath, amsfonts, amssymb, amsthm}
\usepackage{mathtools}
\usepackage{commath}   
\usepackage{hyperref}
\usepackage{color}
\usepackage{url}
\usepackage{float}
\usepackage{paralist}
\usepackage[scr=rsfso]{mathalfa}
\usepackage{amsmath}
\usepackage{hhline}
\usepackage{multirow}
\usepackage{multicol}
\usepackage{tabularx}
\usepackage{comment}
\usepackage{subcaption}
\usepackage{enumitem}
\usepackage{algorithm}
\usepackage{algorithmic}
\usepackage{tabularray}
\usepackage{nicefrac,xfrac}
\usepackage{lmodern}

\begin{document}


\title{Outlier Removal in Cryo-EM via Radial Profiles}
\date{}
\author{Lev Kapnulin, Ayelet Heimowitz, Nir Sharon}
\maketitle
\begin{abstract}
   The process of particle picking, a crucial step in cryo-electron microscopy (cryo-EM) image analysis, often encounters challenges due to outliers, leading to inaccuracies in downstream processing. In response to this challenge, this research introduces an additional automated step to reduce the number of outliers identified by the particle picker. The proposed method enhances both the accuracy and efficiency of particle picking, thereby reducing the overall running time and the necessity for expert intervention in the process. Experimental results demonstrate the effectiveness of the proposed approach in mitigating outlier inclusion and its potential to enhance cryo-EM data analysis pipelines significantly. This work contributes to the ongoing advancement of automated cryo-EM image processing methods, offering novel insights and solutions to challenges in structural biology research.
\end{abstract}

\section{Introduction}\label{sec:intro}

Single-particle cryo-electron microscopy (cryo-EM) is a powerful technique for elucidating high-resolution protein structures in their near-native states. In cryo-EM, a frozen-hydrated biological specimen is imaged in an electron microscope, resulting in experimental images called micrographs. Each micrograph consists of numerous projections of the molecule of interest, as well as contaminants and high levels of noise. Due to the low signal-to-noise ratio (SNR) typically exhibited by micrographs, many particles are required to achieve a 3D protein reconstruction at near-atomic resolution~\cite{cheng2015primer}.  

The process of identifying the individual tomographic projections contained in each micrograph is known as particle picking. While a substantial body of work addresses this problem through both semi-automatic and automatic methods \cite{bepler2019positive,heimowitz2018apple,wagner2019sphire,wang2016deeppicker}, issues such as low SNR, low contrast, crystallized ice, \textit{etc.} still pose substantial impediments to developing precise detection algorithms. Consequently, the output of any particle-picking framework will typically contain a significant amount of outliers. These outliers include overlapping or incomplete particles, carbon, ice crystals, random noise, etc. Using outliers in the reconstruction process can introduce errors and artifacts that adversely impact the quality of the reconstruction~\cite{FRANK1996xv}. As a result, outlier detection is a crucial step in the cryo-EM computational pipeline. 

\begin{figure*}[ht]
    \centering
    \begin{subfigure}[t]{0.75\linewidth}
      \includegraphics[width=\textwidth]{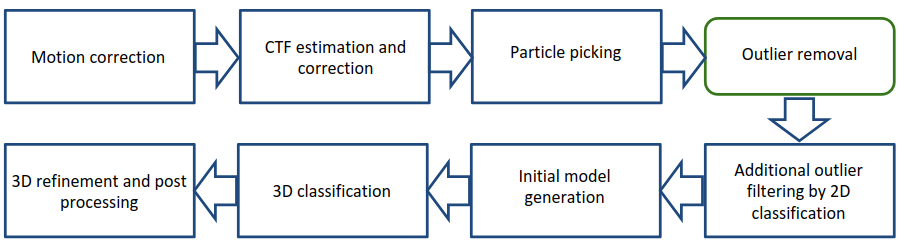}
    \end{subfigure} 
    \caption{Outlier Removal placement in cryo-EM flow.} 
    \label{fig:flow_chart}
\end{figure*}

Several complementary approaches to outlier detection in cryo-EM exist. The most popular approach is known as 2D classification~\cite{singer2020computational,zivanov2018new}. In this approach, similar projection images are identified, aligned, and averaged into class averages. A class average made from true tomographic projections of the macromolecule of interest should contain detail and a much lower level of noise than the individual projections. A class average that includes outlier images will contain noise, a partial projection, overlapping projections, or artifacts stemming from inaccurate correction of the effect of the contrast transfer function (CTF). Expert intervention is then required to identify the ``good'' class averages.

Another notable approach to outlier detection in cryo-EM employs multiple particle pickers concurrently and generates a smart consensus via deep neural network~\cite{sanchez2018deep}. Alternatively, a straightforward pruning that treats the output of the particle-picking algorithm as a mixture of Gaussians has been suggested~\cite{zhou2020unsupervised}. Additionally, another approach uses probabilistic PCA to learn a nonlinear model of the union of subspaces~\cite{weiss2023unsupervised}. 

In this paper, we present a novel, fully automatic framework for outlier detection which addresses three primary categories of outliers: \begin{inparaenum}[(i)]
\item clusters of closely positioned particles,
\item incomplete or partial particles, and
\item erroneously selected noise.
\end{inparaenum}  
Our method is based on the observation that true tomographic projections are bounded by the size of the macromolecule and thus there is a minimal and maximal radius in which its radial profile will remain high.\footnote{We note that in reality, the radius is not determined solely by the size of the macromolecule, as the CTF will increase the size of the projections. We, therefore, slightly increase the maximal radius.} Analysis of such profiles allows for a method of discarding the aforementioned outlier categories.

As such, we provide the mathematical analysis that motivates our method and describe the resulting algorithm in detail. Furthermore, we validate our approach through experiments on both synthetic as well as real-world cryo-EM datasets. Our results demonstrate the proposed framework's capability to maintain high-quality data for reconstruction, streamline the workflow, and improve the accuracy and quality of the final structural models. 

An overview of the computational pipeline within this framework is presented Figure~\ref{fig:flow_chart}. We note that our method is meant as a complementary step to 2D-classification. Unfortunately, in the 2D-classification method, high SNR class averages will be assigned some outliers. These outliers are invisible in the class averages and will be missed by the human expert. Our method, on the other hand, is fully automatic and based on a mathematical analysis of three types of outliers. 2D-classification is still necessary to identify outliers stemming from errors in CTF estimation or artifact. Our method is therefore meant to assist, simplify and speed-up the 2D classification process.

The paper is organized as follows. Section~\ref{sec:our_method} provides a detailed discussion of our method, including a mathematical description. In Section~\ref{sec:numerics}, we present numerical illustrations over synthetic data and real-world datasets. We summarized our findings in Section~\ref{sec:summary}.

\section{Outlier removal in cryo-EM} \label{sec:our_method}

In cryo-EM, projection images that can be considered outliers may arise from various sources, ranging from aggregates or clusters of molecules to ice crystals and carbon film to variability in shot noise~\cite{FRANK1996xv}. We consider any projection that was selected from the micrographs during the particle picking stage yet compromises the accuracy and reliability of the 3D structural determination to be an outlier. However, in this work, we focus on three types of outliers: clusters of closely positioned particles, incomplete or partial particles, and pure noise. We aim to identify as many such projections as possible before performing 3D reconstruction to achieve high-resolution 3D structures. 

\subsection{Notations}

Let $\nu \colon \mathbb{R}^3 \rightarrow \mathbb{R}$ be the Coulomb potential of the 3D volume (particle). The projection operator $\rho:\mathbb{R}^3 \rightarrow \mathbb{R}^2$ is defined as 
\begin{equation*}
    \rho\nu(x_1 ,x_2) := \int_{-\infty}^{\infty} \nu(x_1, x_2, x_3) dx_3
\end{equation*}

Each particle instance is captured in an unknown rotation denoted by $T_j$. The rotated particle is referred to as $T_j\nu$. The resulting $j$th projection image $I_j$ is formed by tomographically projecting $T_j\nu$. This projection is then filtered by the point spread function $\pi_j$ \cite{MINDELL2003334, rohou2015ctffind4,turovnova2017efficient, zhang2016gctf}. Additionally, each projection image is contaminated by high noise levels, primarily resulting from limitations on the number of imaging electrons.

Mathematically, the image formation model of $I_j$ reads, under the weak phase object approximation, as \cite{bhamre2016denoising, FRANK1996xv}:
\begin{equation*}
    I_j = \pi_j * \rho(T_j\nu) + \varepsilon_j,  \quad T_j \in SO(3), \quad j = 1,...,N,
\end{equation*}
where $\varepsilon_j$ is the additive shot noise. Although shot noise follows a Poisson distribution, it is often modeled as an additive, independent, identically distributed (iid) Gaussian noise $\varepsilon_j$. Following the application of phase flipping for CTF correction, we model images selected by a particle picker as:
\begin{equation}\label{eq:base_model}
    I_j = \tilde{\rho}(T_j\nu) + e_j
\end{equation}
We note that where $\varepsilon_j$ follows a Gaussian distribution, so does $e_j$ since phase flipping preserves noise statistics. Additionally, as CTF correction cannot fully restore the original tomographic projections, we denote the resulting tomographic projection as $\tilde{\rho}(T_j\nu)$. As mentioned above, one of the outliers we consider is an image containing multiple closely-positioned particles. To address this issue we modified (\ref{eq:base_model}) as 

\begin{equation}\label{eq:mod_model}
    I_j = \sum_{j_i}\tilde{\rho}(T_{j_i}\nu) + e_{j}.
\end{equation}

\subsection{Template matching} \label{sec:centering}

Template matching is a classical method for object detection in the field of computer vision. It compares an image $I \in \mathbb{R}^{d \times d}$ with a template to find all occurrences of the template in the image. To this end, a template $h \in \mathbb{R}^{k \times k}$ representing the object of interest must be chosen or created. Next, the template is compared to overlapping regions of the larger image $I$ using a similarity metric such as cross-correlation. The aim is to find regions where $h$ aligns closely with features in $I$, indicating the presence and location of the template in $I$.  That is, the peak of the cross-correlation between an image and a template indicates the position where the template best matches the image. This peak provides crucial information for alignment correction, allowing the adjustment of the position of the detected object to achieve proper alignment within the image.

\subsubsection{Application to particle picking}
\begin{figure}[ht]
    \centering
    \begin{subfigure}[t]{0.22\linewidth}
      \includegraphics[width=\textwidth]{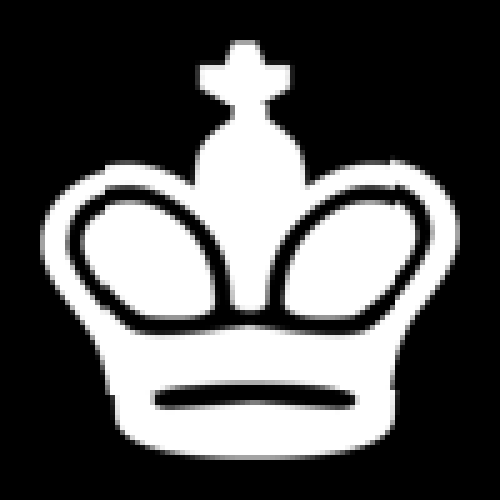}
      \caption{\raggedright A chess piece}
      \label{subfig:chess_piece}
    \end{subfigure}
    \begin{subfigure}[t]{0.22\linewidth}
      \includegraphics[width=\textwidth]{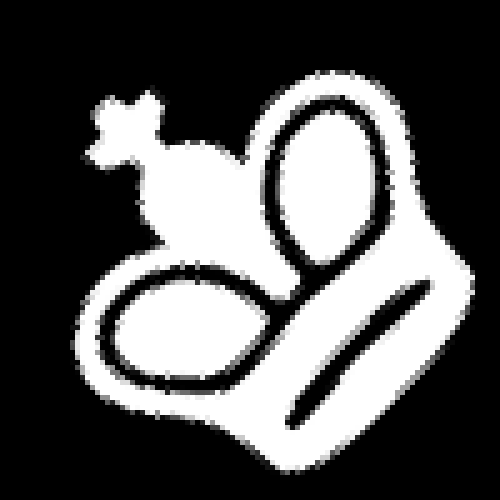}
      \caption*{\raggedright Example 1 of rotated copy}
    \end{subfigure} 
    \begin{subfigure}[t]{0.22\linewidth}
      \includegraphics[width=\textwidth]{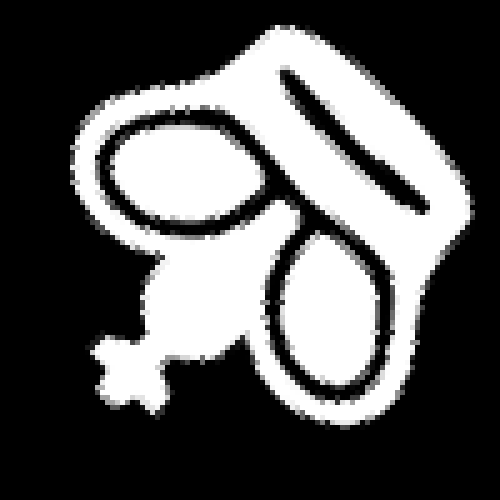}
      \caption*{\raggedright Example 2 of rotated copy}
    \end{subfigure} \quad
    \begin{subfigure}[t]{0.22\linewidth}
      \includegraphics[width=\textwidth]{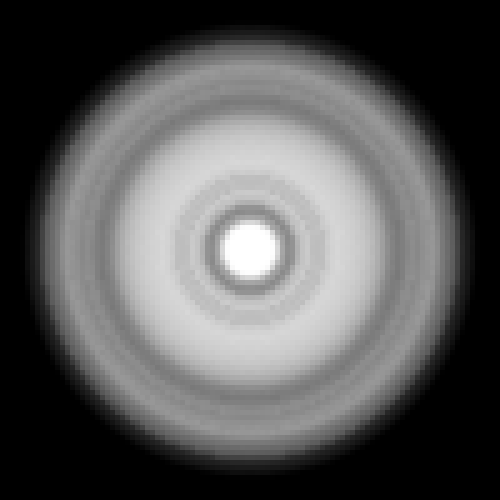}
      \caption{\raggedright The rotational mean around the origin of 2D chess piece.}
      \label{subfig:chess_piece_mean}
    \end{subfigure} 
    \caption{Example of rotational mean. (\subref{subfig:chess_piece}) Rotated copies of a template chess piece. ~(\subref{subfig:chess_piece_mean}) Rotational mean of all possible rotations of our template.}
    \label{fig:rotation_chess_average} \label{fig:rotational_mean}
\end{figure}

In the realm of cryo-EM, obtaining an exact template $h$ that represents the object of interest is quite laborious. For this reason, custom templates, such as mathematical functions, have been employed during particle picking~\cite{vos2009}. Such templates should be invariant to rotations as each particle instance is captured in an unknown orientation. The templates used are, therefore, radially symmetric. One possible template is an approximated radial mean of a 2D tomographic projection rotated around its center of mass. We note that, while the actual template of the projection is difficult to obtain, an approximation of its radial mean is quite straightforward as the mean should contain the highest values in the center, tapering down towards the (known) radius of the projection. An example of such a template is presented in  Figure~\ref{fig:rotation_chess_average}. 

\subsubsection{Application to outlier detection}
We suggest that a radially symmetric, custom mathematical function can be used as a model for the radial profile of the true projections. Specifically, employing a 2D Gaussian template is effective because it approximates a generally spherical shape covering the radial profile of many biological particles' density profiles. 

\begin{figure}[ht]
\centering
    \begin{subfigure}[t]{0.3\linewidth}
      \includegraphics[width=\textwidth]{images/mean_rotation.png}
      \caption*{A rotational mean of an object around the origin}
      \end{subfigure} \quad
    \begin{subfigure}[t]{0.3\linewidth}
      \includegraphics[width=\textwidth]{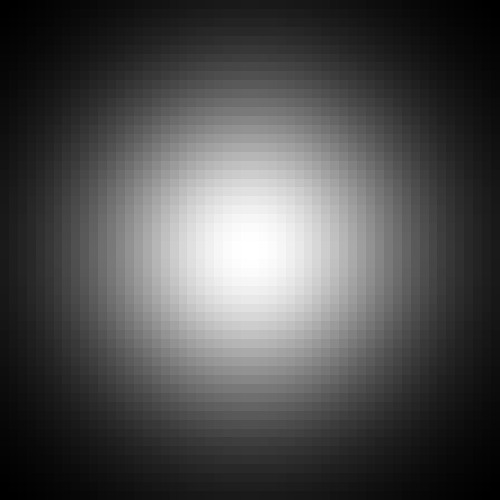}
      \caption*{A sum of 2D Gaussians}
    \end{subfigure} 
    \caption{Comparison of the rotational mean of rotated copies of our chess-piece template from Figure~\ref{fig:rotational_mean} (left) with a sum of Gaussians with different variances, chosen to resemble the same radial support (right).}
    \label{fig:templates}
\end{figure}

In order to increase robustness, we suggest the use of multiple 2D Gaussian templates. The advantage of using multiple templates is that they offer improved accuracy and robustness by accounting for structural variability and different localizations of target particles. An illustration of multiple 2D Gaussian templates and their similarity to rotational mean is illustrated in Figure \ref{fig:templates}.

Furthermore, the application of multiple filters will not incur additional computational costs. This is due to the fact that cross-correlation, as well as convolution, can be computed efficiently in the Fourier space using the fast Fourier transform (FFT). In particular, when using a radially symmetric template $h$, the result of cross-correlation of an image $I$ with $h$ and the result of simple, discrete convolution are identical.

\subsection{Our outlier detector}

The main objective is to detect outliers within the stack of projection images selected in the particle-picking stage. Specifically, our method aims to detect three categories of outliers: clusters of closely positioned projections, incomplete or partial projections, and erroneously selected noise. 

\begin{figure}[ht]
\centering
    \begin{subfigure}[t]{0.3\linewidth}
      \includegraphics[width=\textwidth]{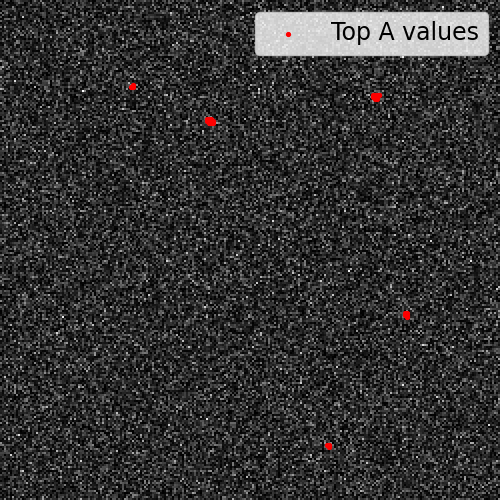}
      \caption{\raggedright Pure noise}
    \end{subfigure} \quad
    \begin{subfigure}[t]{0.3\linewidth}
      \includegraphics[width=\textwidth]{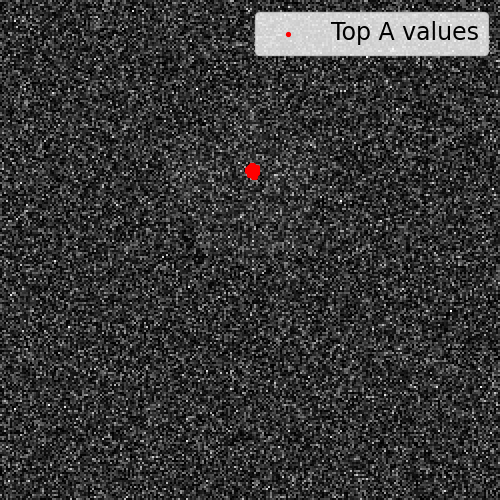}
      \caption{\raggedright Noisy image with SNR=$1/50$}
    \end{subfigure} 
    \caption{Locations of the top $A$ cross-correlation peaks. The left image contains pure noise, and, indeed, the cross-correlation peaks are randomly distributed at multiple locations across the image. In the right image, where a single object is contained, the cross-correlation peaks form a single solid cluster. 
    }
    \label{fig:fig3}
\end{figure}

We apply our method individually to each projection image. We utilize the method outlined in  Section~\ref{sec:centering} to extract the cross-correlation peak of image $I_j$. However, due to high levels of noise, it is possible for pure noise to have a high correlation with our template~\cite{henderson2013avoiding}. Therefore, it is insufficient to rely solely on the single point with the maximum correlation to the template $h$. To increase robustness to noise, we propose extracting the top $A$ cross-correlation peaks. Once these $A$ peaks have been extracted, they can be used to identify outliers. Specifically, a projection matching our template will result in a cluster of peaks densely concentrated in the projection and perhaps some solitary peaks distributed around the image. If the $A$ top peaks form several distinct clusters, we will identify an outlier image containing multiple nearby projections. Where no clusters exist, we identify an outlier of pure noise as images characterized solely by noise typically present a random distribution of the locations of the top $A$ values. We demonstrate this observation in Figure~\ref{fig:fig3}. We, therefore, employ the DBSCAN algorithm~\cite{ester1996density} on the $A$ peaks extracted from each picked particle image. 

\begin{figure}[ht]
\centering
    \begin{subfigure}[t]{0.3\linewidth}
      \includegraphics[width=\textwidth]{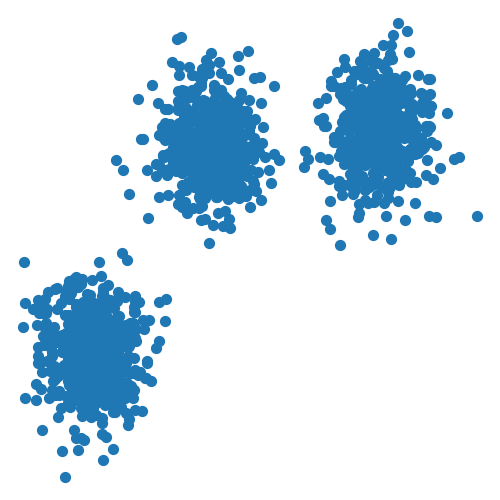}
      \caption*{Sampled data}
      \end{subfigure} \quad
    \begin{subfigure}[t]{0.3\linewidth}
      \includegraphics[width=\textwidth]{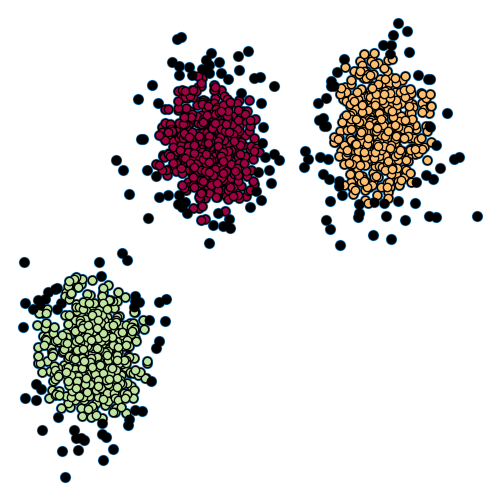}
      \caption*{Clustered data}
    \end{subfigure} 
    \caption{Example of the DBSCAN clustering algorithm. The left plot shows the initial distribution of the sampled data, illustrating three distinct clusters. The right plot displays the clustering results after applying the DBSCAN algorithm with $\varepsilon=0.4$ and $\omega=10$. Each color represents a different cluster identified by DBSCAN, while black points indicate anomalies $\xi$. This figure demonstrates the effectiveness of DBSCAN in identifying clusters of arbitrary shape and size and distinguishing anomaly points $\xi$ in the dataset.}
    \label{fig:dbscan}
\end{figure}

The choice of the DBSCAN algorithm stems from the uncertainty about the exact number of particles in the image. DBSCAN is known for its ability to group data points based on their density distribution without requiring the specification of the number of clusters. We only need to define the standard deviation threshold $\varepsilon$ and the minimum number of points $\omega$ required for a single cluster. It is important to mention that any portion of the $A$ cross-correlation peaks may not belong to any of the recovered clusters. These points are called anomalies $\xi$. An example of the DBSCAN results is presented in Figure~\ref{fig:dbscan}.

\begin{figure}[ht!]
    \centering
    \begin{subfigure}[t]{0.35\linewidth}
      \includegraphics[width=\textwidth]{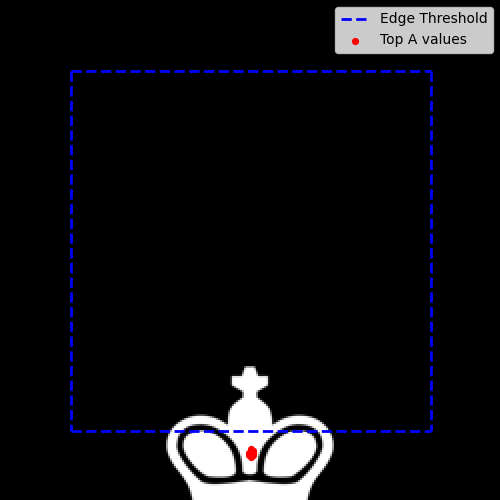}
    \end{subfigure} \qquad
    \caption{Example of a partially contained object. A chess piece where the top $A$ peaks lay outside the threshold $T_E$ indicates an incomplete object.}
    \label{fig:queen_piece}
\end{figure}

To discard outlier images containing partial projections, we identify projection images 
where the center of any cluster of peaks is too close to the border of the image (see Figure~\ref{fig:queen_piece}). We denote the minimal allowed distance from the border as $T_E$ and set $T_E$ to be at least equal to the radius $r$ of the object. Therefore, in images where a single cluster of peaks exists, yet this cluster is close to the image's border, we identify a partial projection and discard such an image. 

\subsubsection{Method summary}

Next, we provide a detailed description of our method, as presented in Algorithm~\ref{alg:outlier}. To initiate the outlier procedure, we require the following parameters: image size $d \times d$,  approximate particle radius $r$, and the number of the peak cross-correlation values to be used $A$. We note that, as shown in the next section, a good initialization of $A$ is approximately $d/5$. Additional parameters are derived from these values: filter size $k = 0.3d$, $h$ is a set of five Gaussian functions with zero mean and a standard deviation that follows the values: $5\%$, $15\%$, $25\%$, $35\%$, and $45\%$ of $k$, edge threshold $T_E = d-r$, DBSCAN local neighborhood threshold $\varepsilon=r$, and DBSCAN minimum number of points $\omega = \frac{A}{3}$ to form a cluster $C$. Once we obtain the $A$ points from the image, we apply DBSCAN to partition them into clusters.

\begin{algorithm}[ht]
\caption{One image filtering}
\label{alg:outlier}
\begin{algorithmic}[1]
\REQUIRE Cropped image $I$ of size $d \times d$, particle radius r and number of highest values A.
\ENSURE Is it an outlier or a valid projection
\STATE $k = 0.3d$
\STATE $T_E = d-r$
\STATE $\omega = \frac{A}{3}$
\STATE $\varepsilon = r$
\FOR{j=0,\ldots, 5}
\STATE $\sigma_j = 0.05 + 0.1 j$
\STATE $h_j$ is a Gaussian of $\sigma_j$ standard deviation
\ENDFOR
\STATE $h = \sum_j h_j$
\STATE $P^\ast = \text{Top}_A(I \ast h)$
\STATE{$C, \xi = \operatorname{Cluster}(P^\ast, \varepsilon, \omega$)}
\STATE{$\xi$ are points which does not belong to any cluster}
\IF{$\#\xi > \omega$}
\RETURN ``Outlier'' \COMMENT{Pure noise}
\ELSIF{$\exists \xi_j > T_E, \quad j =1,..,\#\xi$}
\RETURN ``Outlier'' \COMMENT{Chopped particle}
\ELSIF{$\exists M_{C_j} > T_E, \quad j =1,..,\#C$}
\RETURN ``Outlier'' \COMMENT{Chopped particle}
\ENDIF
\RETURN ``Valid''
\end{algorithmic}
\end{algorithm}

\begin{figure*}[ht]
    \centering
    \begin{tblr}{
   colspec = {X[c,h]X[c,m]X[c,m]X[c,m]X[c,m]}, 
    stretch = 0,
    colsep = 1pt,
    rowsep = 2pt,
    hlines = {1pt},
    }
          {}      & Valid Particle & Pure Noise & Incomplete Particle & Multiple Particles \\
     Clean Image & \includegraphics[width=.6\linewidth]{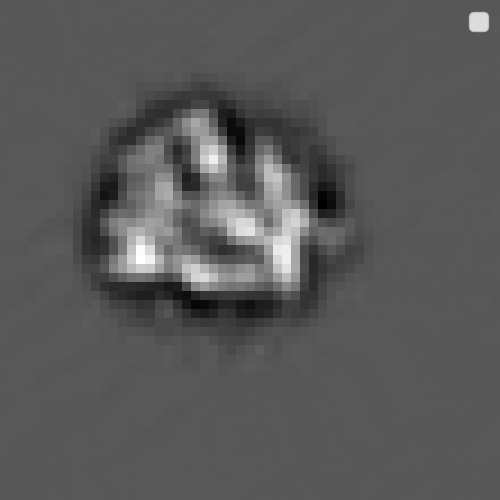} & \includegraphics[width=0.6\linewidth]{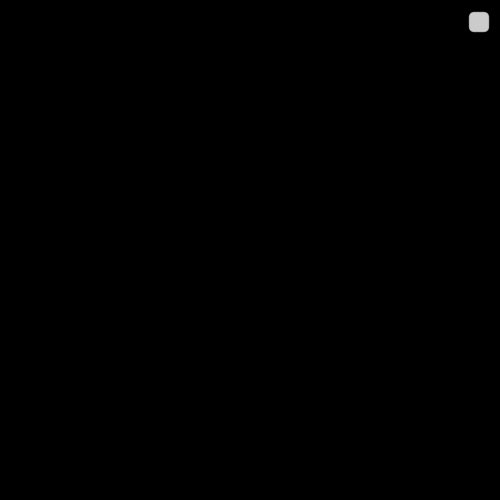} & \includegraphics[width=0.6\linewidth]{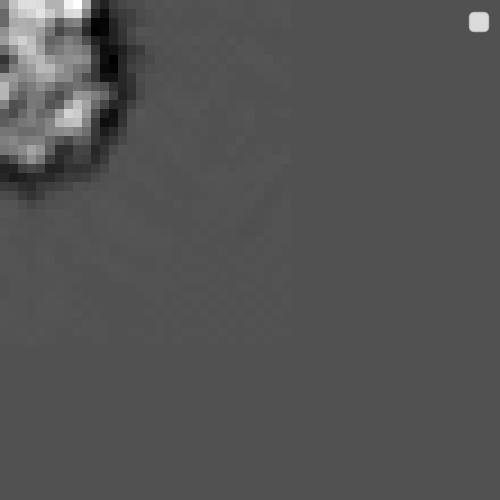} & \includegraphics[width=0.6\linewidth]{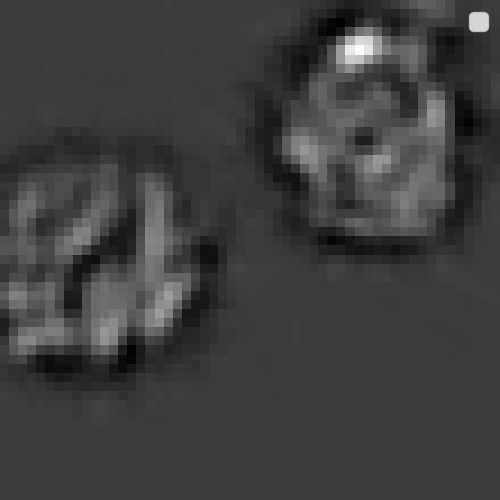} \\
    Noisy Image & \includegraphics[width=0.6\linewidth]{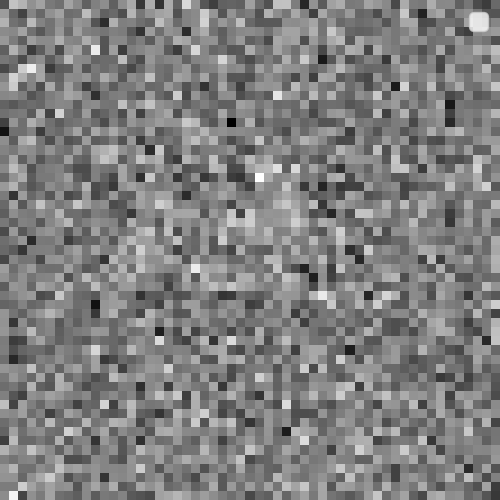} & \includegraphics[width=0.6\linewidth]{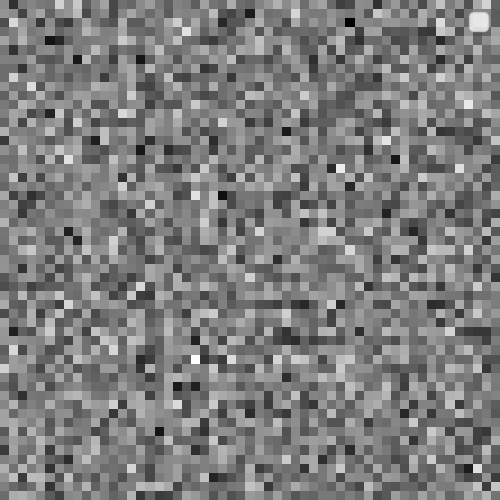} & \includegraphics[width=0.6\linewidth]{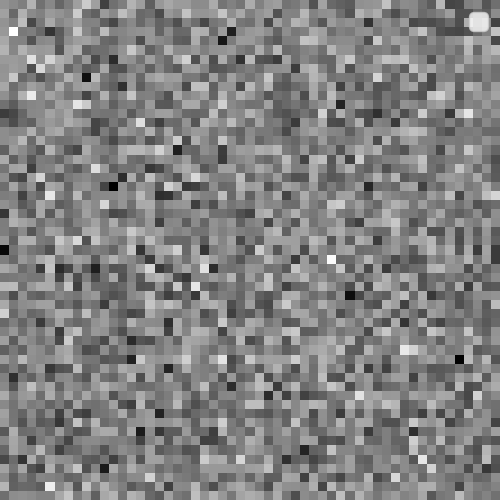} & \includegraphics[width=0.6\linewidth]{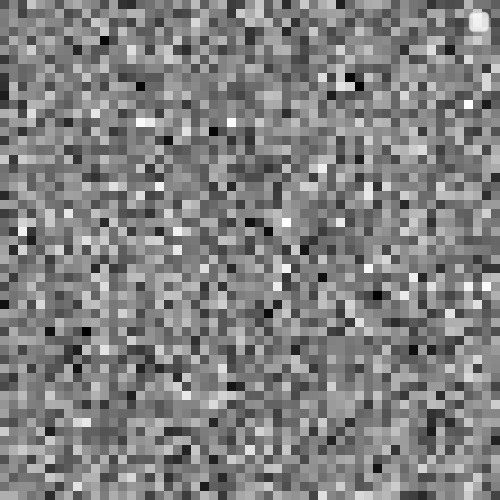} \\
    Cross-Correlation with Marked Peaks & \includegraphics[width=0.8\linewidth]{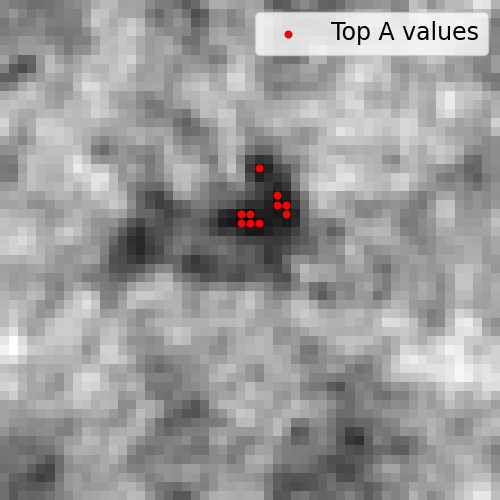} & \includegraphics[width=0.8\linewidth]{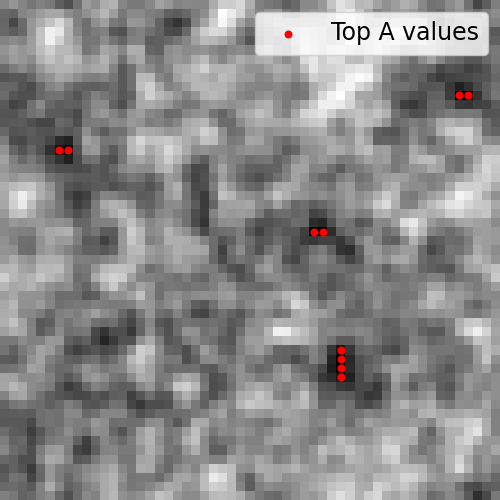} & \includegraphics[width=0.8\linewidth]{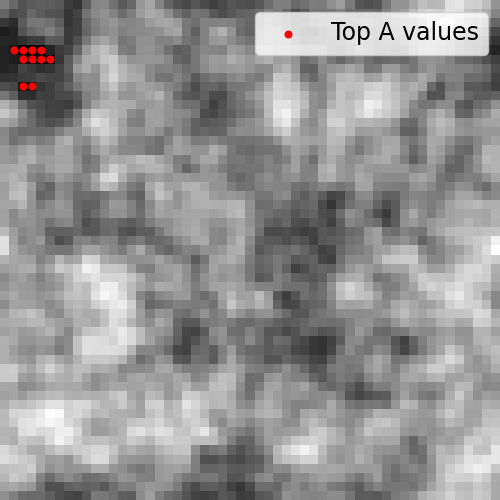} & \includegraphics[width=0.8\linewidth]{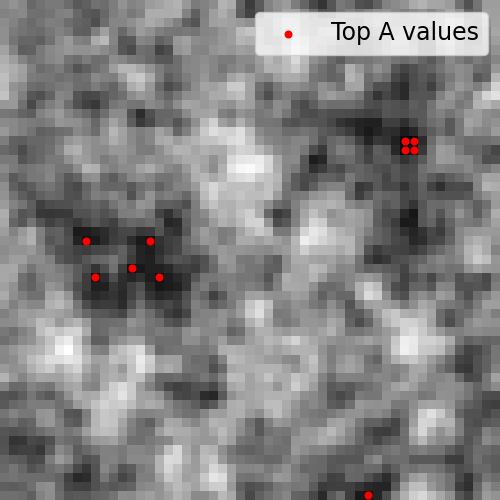} \\
    \end{tblr}
    \caption{Different scenarios of particle detection. The four columns represent, from left to right, the cases of: valid particle, pure noise, incomplete particle, and multiple particles. For each case, we present its clean version, noisy version, and the top $A$ cross-correlation peaks. In the bottom row, the leftmost image of the valid particle shows a single valid cluster located in the center; the second image from the left, representing pure noise, shows multiple randomly distributed peak locations; the second image from the right, showing an incomplete particle, forms a single valid cluster located at the edge; and the rightmost image, representing two particles, shows two valid clusters. The SNR of noised images is approximately $1/80$.}
    \label{tab:synt_images}
\end{figure*}

Once we extract the clusters $C$ and anomalies $\xi$, the initial step involves checking if  $\#\xi > \omega$. If this condition is met, the image is labeled as an outlier. The primary purpose of this condition is to exclude images containing pure noise since, as previously mentioned, pure noise typically induces randomly distributed locations of cross-correlation peaks.

Next, we address incomplete particle cases. Since images with incomplete particles are more likely to have a center $M_{C_i}$ or anomaly $\xi_i$  lying within the threshold $T_E$,  we classify the image as an outlier if at least one of the following conditions is met:
\begin{enumerate}[noitemsep, nolistsep]
    \item $\exists M_{C_i} > T_E, \, \forall i = 1,...,\#C$
    \item $\exists \xi_i > T_E, \, \forall i = 1,...,\#\xi$,
\end{enumerate}
where $M_{C_i}$ is the center of cluster $C_i$. 
The above scenarios are illustrated in Figure~\ref{tab:synt_images}. 
Specifically, Figure~\ref{tab:synt_images} shows four possible scenarios of images picked by particle picker: \begin{inparaenum}[(i)]
\item valid particle,
\item pure noise,
\item incomplete particle, and
\item multiple particles.
\end{inparaenum}
For each scenario, we present the clean image, the image with added noise, and the result of cross-correlation with located top $A$ of the noisy image with the template $h$. 

Our method can be sped up by downsampling the projection images. We note that when the image is downsampled, the parameter $A$, that is, the number of cross-correlation peaks needed to distinguish between the inliers and the different types of outliers, must also be decreased. 

\section{Numerical examples} \label{sec:numerics}

This section presents various simulations where we employ our method on both synthetic and real-world datasets. In Section \ref{sec:synt_data}, we evaluate the performance of our method in controlled environments, each characterized by different outlier types, including one with combined outliers. Subsequently, in Section~\ref{sec:real_data}, we showcase our method's performance over real-world data sourced from the Electron Microscopy Public Image Archive (EMPIAR)~\cite{Iudin2022.10.04.510785}. The complete implementation of our procedure, along with scripts to replicate the experiments on both synthetic and real datasets, is available at \url{https://github.com/lovakap/OutlierRemoval}.

\subsection{Synhtetic Data}\label{sec:synt_data}

This section presents a numerical testing of our method across diverse controlled environments. We aim to demonstrate its robustness and behavior under various scenarios. For the simulation of images selected by a particle picker, we utilized a 3D volume of a reconstructed ribosome 70s~\cite{laurberg2008structural} with size ($65\times65\times65$). First, we define the foundational procedure for image simulation, which will serve as the basis for subsequent simulations. This procedure implements the data model in~\eqref{eq:base_model}, starting with a random rotation applied to the 3D volume and followed by tomographic projection of the rotated volume into 2D. 
This procedure generates inlier (projection) images. To simulate outlier images representing incomplete particles, we apply random translation of size taken in $ \left[\frac{d}{4}, \frac{2d}{3}\right]$ and with a random plane rotation by an angle from $\left( 0, 2\pi\right]$, where $d$ represents the image size\footnote{We assume the image size satisfies $d<2r$ and the random translation is selected to ensure that part of the object lies outside the image boundaries}. Both translation and rotation are sampled from a uniform distribution, guaranteeing that, at most, only half of the particle will be within the image but never less than a third of the particle.

\begin{figure}[ht]
    \centering
    \begin{subfigure}[t]{0.45\linewidth}
      \includegraphics[width=\textwidth]{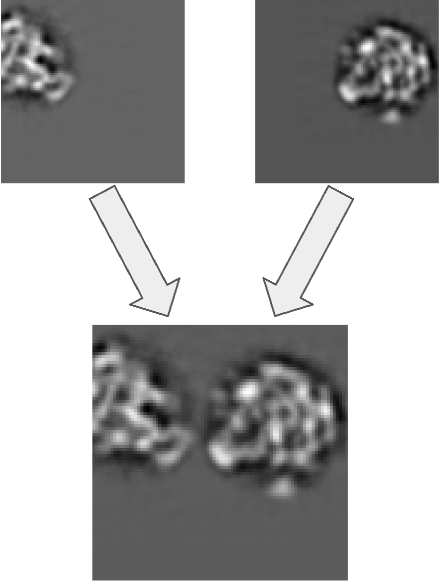}
    \end{subfigure} 
    \caption{An illustration of an image containing multiple particles. We combine the images by taking their mean value to maintain consistent range values.} 
    \label{fig:multi_comb}
\end{figure}

Projection images containing multiple particles were created by adding two images of partial particles. However, for these images we use a translation in $\left[\frac{d}{10}, \frac{2d}{3}\right]$. This adjustment enabled the generation of images containing both partial and complete particles. This process is illustrated in Figure~\ref{fig:multi_comb}.

Following the creation of inlier and outlier images, we contaminate the projection images with colored noise. This noise exhibits a spectral decay resembling $\sfrac{1}{\sqrt{1 + \rho^2}}$ along the radial direction $\rho$ \cite{heimowitz2021}. In our experiments, the noise level is quantified in terms of signal-to-noise ratio, which we expressed as SNR $ = \sfrac{\mu}{\sigma}$, where $\sigma$ represents the standard deviation of the noise, and $\mu$ signifies the mean value of the clean image. We specifically calculated the mean value within the vicinity of the object, as opposed to considering the entire image. In the case of pure noise, we cannot compute SNR for an empty image; therefore, to maintain consistency in the calculation, we exclusively compute the mean SNR value for the inlier images as the overall representative SNR.

We conducted an experiment on synthetic data using a dataset of $2000$ images, where $1000$ images were inliers and the remaining were outliers. This experiment assessed a combination of all possible outlier types, including pure noise, incomplete particles, and multiple particles. Subsequently, we applied our method to detect the outliers. Table~\ref{tab:synt_images} shows examples of valid particles and each possible outlier.

\begin{figure}[ht]
    \centering
    \begin{subfigure}[t]{0.4\linewidth}
      \includegraphics[width=\textwidth]{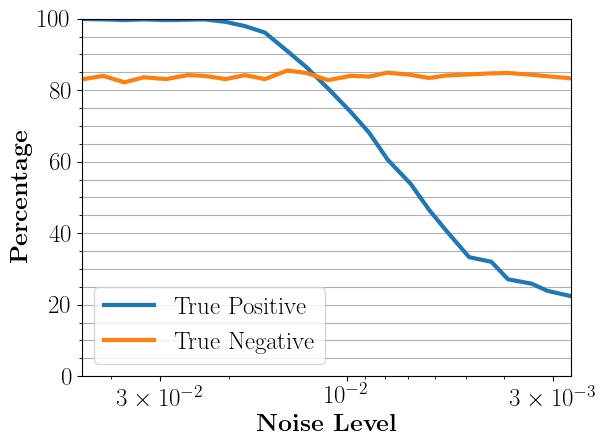}
    \end{subfigure} 
    \caption{Performance of the algorithm at different noise levels on combined outliers, which is expressed in terms of SNR. The X-axis is in inverse order and on a logarithmic scale.} 
    \label{fig:sim_res}
\end{figure}

In this experiment, we used a particle image size of $d=33$, radius $r = 5$ and $A=10$. We note that each experiment was repeated for various levels of noise. We evaluated our performance using the following metrics: 
\begin{compactenum} 
    \item True positive rate (TPR): \% of inlier images that correctly classified as such.
    \item True negative rate (TNR): \% of successfully detected outliers,
\end{compactenum}
A summary of the results is presented in Figure~\ref{fig:sim_res}. It is evident that we successfully retained 90\% to 100\% of the inlier images while removing about 85\% of outliers, even at extremely low SNRs such as 1/70. Additionally, we conducted similar experiments where we assessed each type of outlier individually. The results of these experiments were consistent with the combined outlier experiment, further validating our method.

\begin{figure}[ht]
    \centering
    \begin{subfigure}[t]{0.35\linewidth}
      \includegraphics[width=\textwidth]{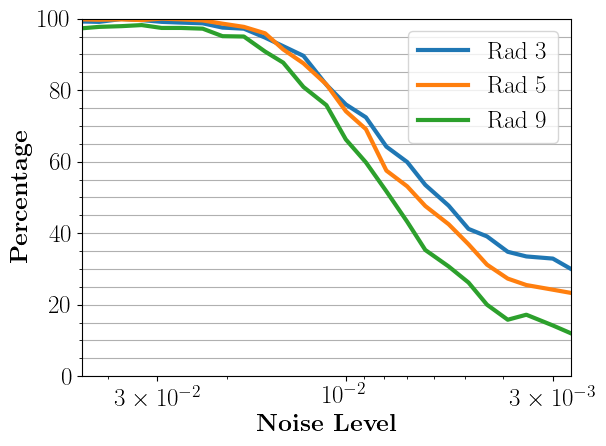}
       \caption*{True positive rate}
    \end{subfigure} 
    \begin{subfigure}[t]{0.35\linewidth}
      \includegraphics[width=\textwidth]{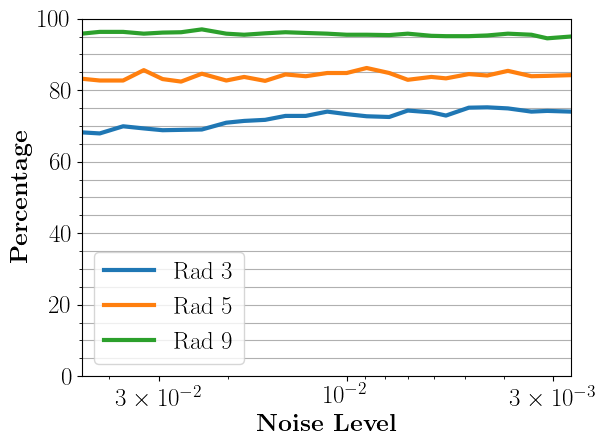}
      \caption*{True negative rate}
    \end{subfigure} 
    \caption{True positive and true negative rates for $r=3,5,9$.} 
    \label{fig:combined_graphs}
\end{figure}

Lastly, we note that the radius of the particle $r$, provides our method with a degree of control over the strictness of the selection criteria. By reducing the size of $r$, we can pick more inlier images at the expense of including additional outlier images, and vice versa.  We demonstrate this flexibility in Figure~\ref{fig:combined_graphs}. We note that in the context of cryo-EM, the error of false positives is less detrimental than that of false negatives, as the imaging process creates a bottleneck. We therefore recommend the use of smaller values of $r$.

\subsection{Real Data}\label{sec:real_data}

Next, we demonstrate our suggested method using real-world cryo-EM datasets. Specifically, we conducted tests on two structures: 80S ribosome and $\beta$-galactosidase. The corresponding datasets are publicly available in the Electron Microscopy Public Image Archive (EMPIAR) as 10028 \cite{wong2014cryo} and 10017 \cite{scheres2015semi}, respectively.

To handle the outliers identified by the particle picker, we developed a procedure that integrated with the full cryo-EM reconstruction workflow. We used Relion 4.1~\cite{kimanius2021new} for this workflow, which included movie correction, CTF correction (using the built-in CTFFIND4~\cite{rohou2015ctffind4}), particle extraction, 2D classification, 3D classification, and final reconstruction which includes the stages of 3D refinement and post-processing. Our method requires CTF-corrected extracted particles, so we integrated our method immediately after particle extraction.
In our Python implementation, we utilized Python-Aspire~\cite{wright2022computationalcryoem} to manage the extracted and CTF-corrected particles.

\subsubsection{80S ribosome}

The EMPIAR-10028 dataset consists of 1081 micrographs, each with dimensions of  $4096 \times 4096$, as well as the coordinates of 160,712 projection images. We cropped these images from the micrographs using a window size of $280\times280$. To improve efficiency, we down-sampled the images to a size of $d=55$, which significantly reduced the procedure’s runtime. Additionally, we set the parameters $r=17$ and $A=10$. After applying our outlier detection method, we identified 43,860 outlier images. This process took 14.7 minutes to complete on 3.1 GHz Intel Xeon Gold 6254 CPU.

To evaluate the efficacy of the outlier removal method, we conducted Relion's 2D classification on both the original 160,712 projection images and on the 116,852 images identified by our method as inliers. Following the 2D classification, we manually excluded the undesirable classes. In the original dataset, 96,410 particles remained, representing 60\% of the total, while the filtered dataset retained 99,322 particles, or 85\% of the inliers.

It's important to note that due to the reduced size of the filtered dataset, the running time of the 2D classification improved by 16.1 minutes, while the number of images after elimination remained roughly the same as the original dataset after the elimination. In addition, after running the final reconstruction on both datasets, we achieve a comparable resolution of 3.71\r{A} on inliers and 3.77\r{A} on the original dataset. The results are summarized in Table \ref{tab:extraction_rate_10028}.

\begin{table}[ht!]
    \centering
    \begin{tabular}{|c||c|c|}
        \hline
        & Original & Ours \\ \hline \hline
        \# coordinates & 160,712 & 116,852 \\ 
        \hline
        2D classification & 96,410 (59\%)   &  99,322 (85\%) \\
        \hline
        3D classification & 96,410 (59\%)   &  99,322 (85\%) \\
        \hline
        Resolution & 3.77\r{A}   &  3.71\r{A} \\
        \hline
        2D Time & 67 min  &  50.9 min \\
        \hline      
    \end{tabular}
    \caption{Comparison of 3D reconstruction for the 10028 dataset.}
    \label{tab:extraction_rate_10028}
\end{table}

\subsubsection{\texorpdfstring{$\beta$}{Beta}-galactosidase}

We applied the same procedure to the EMPIAR-10017 dataset, which comprises 84 micrographs, each with dimensions of $4096 \times 4096$ pixels. For this dataset, we cropped images to $120 \times 120$ pixels and down-sampled them to $d=55$. We also adjusted key parameters to $r=8$ and $A=10$. Out of a total of 40,863 images, our outlier detection process identified 5,840 outliers, leaving us with 35,023 inlier images. The entire process was completed in 50 seconds using the same computational setup.

Since the particles in this dataset were manually selected by experts, we anticipated that our method’s performance might be lower compared to datasets curated by semi- or fully-automated particle pickers. However, the smaller dataset size resulted in a 5.5-minute reduction in the 2D classification time. After completing the 2D classification, we manually excluded undesirable classes from both the original and inliers datasets, leaving us with 35,201 and 34,418 images, respectively. Both datasets ultimately achieved the same 3D reconstruction resolution of 4.36\r{A} after removing the undesirable classes identified during 2D and 3D classification.

\begin{table}[ht!]
    \centering
    \begin{tabular}{|c||c|c|}
        \hline
        & Original & Ours \\ \hline \hline
        \# coordinates & 40,863 & 35,023 \\ 
        \hline
        2D classification & 35,201 (86\%)   &  34,418 (98\%) \\
        \hline
        3D classification & 29,814 (72\%)   &  26,292 (75\%) \\
        \hline 
        Resolution & 4.36\r{A}   &  4.36\r{A} \\
        \hline
        2D Time & 39.3 min  &  33.7 min \\
        \hline       
    \end{tabular}
    \caption{Comparison of 3D reconstruction for the 10017 dataset.}
    \label{tab:extraction_rate_10017}
\end{table}

\begin{figure}[ht]
    \centering
    \begin{subfigure}[t]{0.3\linewidth}
      \includegraphics[width=\textwidth]{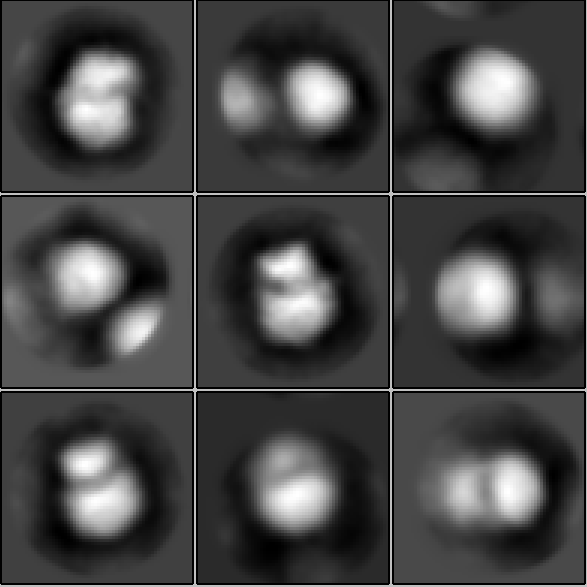}
      \caption*{\raggedright 2D class averages of detected outliers}
    \end{subfigure}  \quad
    \begin{subfigure}[t]{0.3\linewidth}
      \includegraphics[width=\textwidth]{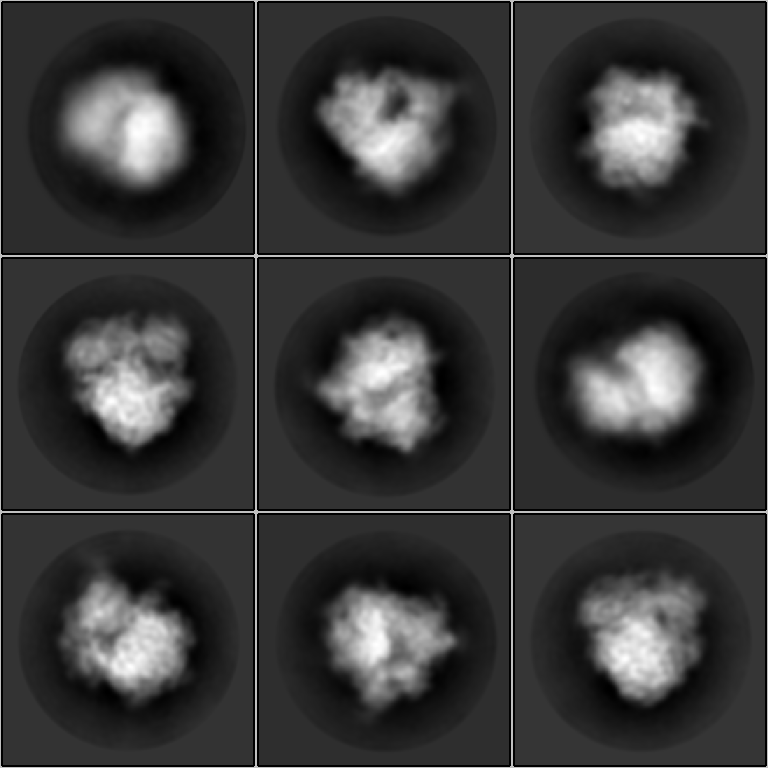}
      \caption*{ \raggedright 2D class averages of inlier images post-outlier removal.}
    \end{subfigure}
    \caption{Comparison of 2D class averages from the EMPIAR-10028 dataset. The left plot shows the sampled 2D class averages of outlier images identified by our procedure, illustrating their variability and lower quality. The right plot displays the sampled 2D class averages of inlier images after removing outliers, demonstrating the improved consistency and quality of the remaining data.} 
    \label{fig:10028_classes}
\end{figure}

\begin{figure}[ht]
    \centering
    \begin{subfigure}[t]{0.3\linewidth}
      \includegraphics[width=\textwidth]{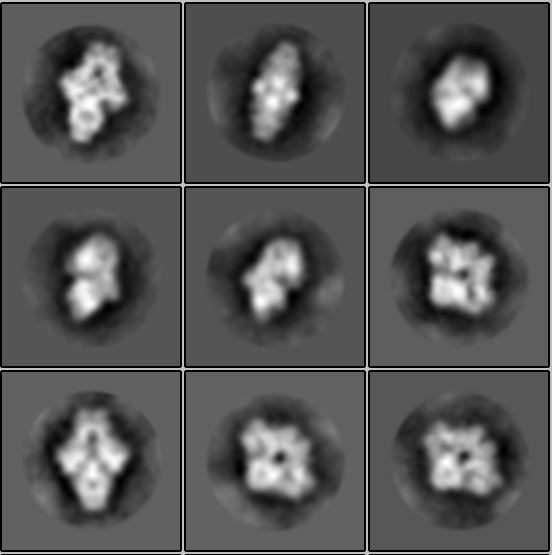}
      \caption*{\raggedright 2D class averages of detected outliers}
    \end{subfigure}  \quad
    \begin{subfigure}[t]{0.3\linewidth}
      \includegraphics[width=\textwidth]{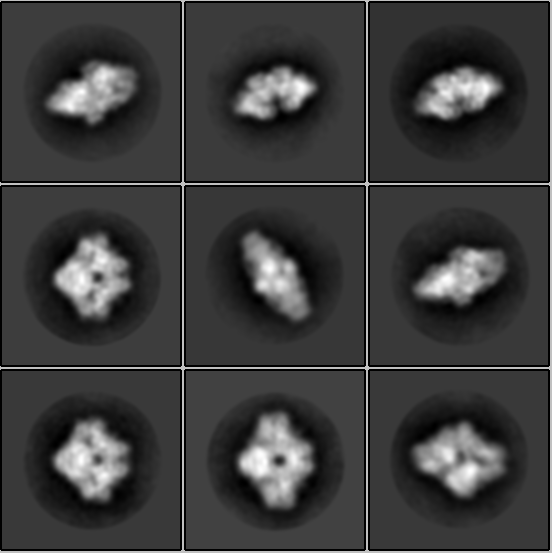}
      \caption*{ \raggedright 2D class averages of inlier images post-outlier removal}
    \end{subfigure}
    \caption{Comparison of 2D class averages from the EMPIAR-10017 dataset. The left plot shows the sampled 2D class averages of outlier images identified by our procedure, illustrating their variability and lower quality. The right plot displays the sampled 2D class averages of inlier images after removing outliers, demonstrating the improved consistency and quality of the remaining data. Classes are arranged in descending order based on their distribution frequencies.} 
    \label{fig:10017_classes}
\end{figure}

In Figures \ref{fig:10028_classes} and \ref{fig:10017_classes}, it is evident that the images we discarded do not form high-quality class averages, as shown in the left plots. The variability and lower quality of these outliers contrast with the improved consistency and quality of the class averages depicted in the right plots, where the inlier images after outlier removal are presented.

\section{Summary} \label{sec:summary}

Our work addresses the challenge of identifying outlier images picked by the particle picker, which can significantly hamper the reconstruction process. We introduced an automatic procedure for outlier removal that effectively removes these outliers. Our approach is based on detecting pixels with high correlation to a predetermined sum of Gaussians, allowing for more accurate identification and exclusion of outlier images. By implementing this procedure, we reduced the time required for the reconstruction process without damaging the final resolution of the resulting images.

We thoroughly evaluated the performance of our outlier removal procedure using both synthetic data and actual cryo-EM data. For the synthetic data, we tested various levels of SNR to ensure robustness and effectiveness under different conditions. Our results consistently demonstrated that the procedure could efficiently filter out noise and irrelevant particles, maintaining high-quality data for reconstruction. When applied to real cryo-EM datasets, the procedure streamlined the workflow while maintaining the accuracy and quality of the final structural models.

\clearpage
\pagebreak

\section*{Acknowledgement}

NS is partially supported by the DFG award 514588180 and the NSF-BSF award 2019752. 
 
 The authors acknowledge the Ariel HPC Center at Ariel University for providing computing resources that have contributed to the research results reported in this paper

\bibliographystyle{plain}
\bibliography{main}

\end{document}